# Top Quark Physics at the Tevatron[a]

C.–P. Yuan

*Department of Physics and Astronomy*
*East Lansing, MI 48824 , USA*

We discuss physics of the top quark at the Fermilab Tevatron. By the year 2000, many properties of the top quark can be measured at the Tevatron.

## 1  Discovery of the Top Quark

The discovery of the top quark is one of the most important discoveries at the Fermilab Tevatron which is currently the only collider that produces top quark on its mass-shell. The mass of the top quark has been measured to be $m_t = 176 \pm 8\,(\text{stat.}) \pm 10(\text{sys.})\,\text{GeV}$ by CDF group and $m_t = 199^{+29}_{-21}\,(\text{stat.}) \pm 22(\text{sys.})\,\text{GeV}$ by DØ group, by detecting the $t\bar{t}$ events. The standard $t\bar{t}$ event selection is based on the expected Standard Model (SM) decay chain $t\bar{t} \to (W^+ b)(W^- \bar{b})$ and the subsequent decays of the $W$'s into fermion pairs. At least one $W$ is tagged in the mode $W \to \ell + \nu$ by requiring an isolated high $p_T$ (transverse momentum) lepton ($\ell = e$ or $\mu$) and large $\not{E}_T$ (missing transverse energy). In the "dilepton" analysis the leptonic decay of the other $W$ is identified with a loose lepton selection; this mode has small backgrounds but small branching ratio of just $4/81 \simeq 5\%$. In the case of the "lepton + jets" mode, the second $W$ decays to quark pairs, giving larger branching ratio of $24/81 \simeq 30\%$. The final state of $(\ell \nu b)(jj\bar{b})$ is separated from the primary background, $W +$ jets, by requiring a large multiplicity of high $p_T$ jets and also evidence of a $b$-jet, using either secondary vertex (silicon detector) or soft lepton ($b \to c \ell \nu X$) identification. For more detailed discussions on the event selection and the detector configuration which determines the acceptance and the detection efficiency of the events, we refer the readers to Ref. [1].

---

[a] Talk given at the International Workshop on Elementary Particle Physics: Present and Future, Univ. of Valencia, Valencia, Spain, June 5 to 9, 1995; and at the Fermilab Users Annual Meeting, Fermilab, Illinois, July 13 & 14, 1995.

Assuming the SM decay mode of the top quark, the cross section $\sigma_{t\bar{t}}$ for the QCD production processes $q\bar{q}, gg \to t\bar{t}$ was measured to be $7.6^{+2.4}_{-2.0}$ pb (by CDF) and $6.4 \pm 2.2$ pb (by DØ). For comparison, the SM result for $m_t = 175\,\text{GeV}$ at $\sqrt{s} = 1.8\,\text{TeV}$ is $\sigma_{t\bar{t}} = 5.52^{+0.07}_{-0.45}$ pb quoted from Ref. [2] in which the effects of multiple soft-gluon emissions have been properly resummed. Since the measurement of the cross section (obtained from the "counting" experiments, i.e. counting the observed total $t\bar{t}$ event numbers in various decay modes) and the measurement of the mass of the top quark (obtained from reconstructing the invariant mass of the top quark) are two independent measurements, one can combine these results to find the best fitted values for $m_t$ and $\sigma_{t\bar{t}}$ [3]. We find that $m_t = 174 \pm 4\,\text{GeV}$ and $\sigma_{t\bar{t}} = 5.9 \pm 0.7$ pb [4]. If the $t\bar{t}$ event rate is indeed given by the SM prediction for $\sigma_{t\bar{t}}$, then the results of the fit described above conclude the branching ratio (BR) of $t \to X$ for $X \neq bW$ has to be less than $\sim 10\%$ [4].

## 2 Is Top Quark just another heavy quark?

$m_t$ ($\simeq v/\sqrt{2} = 174\,\text{GeV}$) is about the scale ($v$) of the electroweak symmetry breaking (EWSB). Studying top quark might provide clues to generation of the fermion masses which could be closely related to the EWSB. Furthermore, effects of new physics originating from the EWSB would be more apparent in the top quark sector than any other light sector of the electroweak theory. Hence, the top quark system not only serves as a stage for testing the SM but also provides a window to new physics beyond the SM.

A few examples are discussed in Ref. [5] to illustrate that different models of EWSB mechanism will induce different interactions among the top quark and the $W$- and $Z$-bosons. These interactions may strongly modify the production and/or the decay of the top quark. In some models (e.g., the TopColor model [6]) observable flavor-changing neutral current (FCNC) processes (e.g., $t \to cZ, cg, c\gamma \ldots$) can be mediated by new underlying dynamics, and some new resonances can strongly couple to $t\bar{t}$ (e.g., a degenerate, massive color octet of "colorons" and a singlet heavy $Z'$) or $t\bar{b}$ (e.g., a triplet of "top-pions") system. With all these new effects possibly appearing in the top quark system, we conclude that top quark is likely not just another heavy quark.

Is it a SM top quark? What do we know about the interaction of the top quark? Can we learn about that from the radiative effects to the precision LEP/SLC data (physics at the $Z$-pole)? A few analyses for studying the top quark couplings to the gauge bosons show that current low energy data still allow rooms for new physics [5, 7], and that only the direct measurements (with on-shell top quark produced) of these couplings can be conclusive.

## 3 Top Quark Physics for the Tevatron at Run-II and Beyond

To determine how well an observable can be measured at the Tevatron in the Main Injector Era (Run-II and beyond), we need to set up a reference for top quark event rates. For this purpose, we consider a $\bar{p}p$ collider with $\sqrt{S} = 2$ TeV and an integrated luminosity of $2\,\text{fb}^{-1}$ (or, $1\,\text{fb}^{-1}$ for each experimental group). In the SM, for a 175 GeV top quark, there will be about $1.4 \times 10^4$ $t\bar{t}$ pairs and $5 \times 10^3$ single-$t$ or single-$\bar{t}$ events produced. After taking into account the $b$-tagging efficiency and the detection efficiency [1], there are about 1000 single-$b$-tagged $t\bar{t}$ pairs in the $\ell+$ jets sample (among those 600 are also double-$b$-tagged), 100 in the dilepton sample, and 250 single-$t$ or single-$\bar{t}$ events (in the $\ell+$ jets sample) available for testing various properties of the top quark. In the following, we discuss the relevant observables and show that with a $2\,\text{fb}^{-1}$ luminosity, many *first* measurements can already be done to a good accuracy at the Tevatron. With a $(10-100)\,\text{fb}^{-1}$ integrated luminosity (beyond Run-II), many further improvements are expected.

### 3.1 $t\bar{t}$ production rate $\sigma_{t\bar{t}}$

At the Tevatron, the dominant $t\bar{t}$ pair production mechanism is $q\bar{q} \to t\bar{t}$ not $gg \to t\bar{t}$, the former contributes about 90% of the rate because the quark luminosities are larger than the gluon luminosities for large x (i.e. for producing heavy top quarks). To test QCD, we need an accurate measurement for $\sigma_{t\bar{t}}$ which is experimentally limited by the systematic uncertainties. It is concluded in Ref. [1] that the experimental error in $\sigma_{t\bar{t}}$ is $\delta\sigma_{t\bar{t}} \simeq 10\%$ which is about the same as the theoretical error for calculating $\sigma_{t\bar{t}}$ [2]. (Note that counting the dilepton event rate is a direct measurement of $\sigma_{t\bar{t}}$, and the statistical error for the dilepton sample of $t\bar{t}$ events is about $1/\sqrt{100} = 10\%$.)

### 3.2 Mass of the top quark $m_t$

With enough $t\bar{t}$ pair events, the accuracy in measuring $m_t$ will be determined by the systematic uncertainty which is dominated by the error in measuring the jet energy scale due to the imperfections in the calorimetry and the effects of initial and final state gluon radiations. The determination on the jet energy scale can be greatly improved by studying the $Z+1$ jet and $\gamma+1$ jet events [1]. It is expected that $m_t$ can be measured to within a couple percent. So, the uncertainty in measuring $m_t$ is $\delta m_t \simeq (2-4)\,\text{GeV}$.

### 3.3 Distributions of invariant mass $M_{t\bar{t}}$ and transverse momentum $p_T(t)$

If a heavy new resonance ($V$) can be produced in $\bar{p}p$ collision and can strongly couple to $t\bar{t}$ [8], then the observed distributions of $M_{t\bar{t}}$ and $p_T(t)$ will be different from the SM predictions. (The event rates can either increase or decrease.) By carefully comparing these distributions with the predictions of a given theory, one can then either approve or exclude that theory. Since $M_{t\bar{t}}$ can be reconstructed on an event-by-event basis by requiring that there are two top quarks observed in the event, the shape and the magnitude of this distribution can therefore set a model-independent limit on possible new physics coupled to $t\bar{t}$ pairs. Demanding a resonance to be observed at the $5\sigma$ level (i.e. $S/\sqrt{B} \gtrsim 5$) over the $t\bar{t}$ continuum in the $M_{t\bar{t}}$ spectrum, one can set the minimum bound on $\sigma(\bar{p}p \to V) * (V \to t\bar{t})$ to about $(0.4 - 0.8)$ pb and $(0.1 - 0.2)$ pb for $M_V$ equal to 500 GeV and 800 GeV, respectively [1].

### 3.4 Top quark decays and FCNC decay modes of top

Because top quark is heavy, it will decay via weak interaction before it feels non-perturbative strong interaction. This is the first opportunity we have for studying the properties of a bare quark. In the SM, the total decay width of a SM top quark is $\Gamma_t \simeq 1.6\,\text{GeV}(m_t/180\,\text{GeV})^3$, and the branching ratio of the weak two body decay $t \to bW^+$ is about one hundred percent. In this decay mode the top quark will analyze its own polarization [9]. If $t$ is found dominantly decaying to $bW^+$, the second top ($\bar{t}$) in each $t\bar{t}$ event should be carefully studied as a window for small non-SM decay modes of top quarks. We call this $t$-tagging.

In the SM, the branching ratios for the FCNC decay modes were found to be too small to be detected, but they can be observable in some other models. Consider the MSSM with light chargino and top-squark. One of the $t$ in the $t\bar{t}$ event decays to $\tilde{t}_1 \tilde{\chi}_1^0$, and $\tilde{t}_1$ subsequently decays to $c \tilde{\chi}_1^0$. The signature of this event is $W + 2\,\text{jet} + \slashed{E}_T$ which is not included in the counting experiments that only count events with $W+ \geq 3\,\text{jets}$. A careful study on this signature can approve the MSSM or set limit on the MSSM parameters [4]. Other studies [1] show that a $2\,\text{fb}^{-1}$ luminosity can be sensitive to $\text{BR}(t \to cZ) \sim 2\%$ (from $3\ell + 2\,\text{jets}$ or $2\ell + 4\,\text{jets}$ sample) and $\text{BR}(t \to c\gamma) \sim 0.3\%$ (from $\ell + \gamma + 2\,\text{jets}$ or $\gamma + 4\,\text{jets}$ sample).

### 3.5 Ratios of branching ratios: $\mathcal{R}_\ell$ and $\mathcal{R}_b$

Define $\mathcal{R}_\ell$ to be the ratio of the $t\bar{t}$ cross sections measured using dilepton events to that measured using $\ell +$ jets events. If $\mathcal{R}_\ell$ differs from the SM prediction, then it implies new physics that would allow $t$ decay without a $W$ boson in the final state, such as charged Higgs ($t \to bH^+$ [10]) and top-squark ($t \to \tilde{t}_1 \tilde{\chi}_1^0$). Hence, $\mathcal{R}_\ell$ measures BR($t \to bW$). With $2\,\mathrm{fb}^{-1}$, the error on BR($t \to bW$) is about 10% [1]. Another useful ratio $\mathcal{R}_b$ is the ratio of the $t\bar{t}$ cross sections measured using double-$b$-tagged events to that measured using single-$b$-tagged events. This determines the upper limit on the branching ratio of $t \to X$ where $X$ does not contain any $b$-quark. (This can be applied to both $\ell+$ jets and dilepton samples, from a known $b$-tagging efficiency.) With $2\,\mathrm{fb}^{-1}$, this upper limit can be set to about $(3-5)\%$ [1]. This result can be interpreted as the error on measuring BR($t \to bW$) if a $W$ boson is confirmed in $t$ decay. (Generally a small BR($t \to X$) is better measured from direct search of the rare decay mode than from the measurements of $\mathcal{R}_\ell$ and $\mathcal{R}_b$.)

### 3.6  Partial decay width $\Gamma(t \to bW)$ and the lifetime of top

The total decay width $\Gamma_t$ of a SM top quark cannot be measured from the invariant mass of $t$ reconstructed in the $t\bar{t}$ events [11]. An elegant way to determine the lifetime (the inverse of the total decay width) of the top quark is to measure the partial decay width $\Gamma(t \to bW)$ and the branching ratio BR($t \to bW$), since $\Gamma_t = \Gamma(t \to bW)/\mathrm{BR}(t \to bW)$. As shown in Ref. [11], $\Gamma(t \to bW)$ can be *model-independently* measured by counting the production rate of single-top events produced from the $W$-gluon fusion process (which is *equivalent* to the $W$-$b$ fusion process), because the production rate of $Wb \to t$ is directly proportional to the decay rate of $t \to bW$. The cross section for the $W$-gluon fusion process is known to about $(15-20)\%$ [11], the lifetime of the top quark can therefore be determined to about $(20-30)\%$.

Before concluding this section, we note that measuring the single-top event rate from the $W$-gluon fusion process is an *inclusive* method for detecting effects of new physics which might produce large modifications to the interactions of the top quark [11]. This provides the *first* hint of possible large new physics effects in the production of the single-top events at the Tevatron, and a further detailed test of the $t$-$b$-$W$ couplings can be achieved by studying the decay of top quarks in both the $t\bar{t}$ pairs and the single-top events.

### 3.7  Form factors of $t$-$b$-$W$, $V_{tb}$ and $m_t$

To describe the decay of $t \to bW(\to \ell\nu)$, the most general form factors contain $f_1^{L,R}$ and $f_2^{L,R}$ [9]. Using the invariant mass ($m_{b\ell}$) of $b$ and $\ell$, one can determine the polarization of the $W$ boson which depends on the values of these form factors. A study [12] shows that the errors $\delta f_1^L \sim (2-3)\%$ and $\delta f_1^R \sim 0.2$, if $f_1^L \approx 1$ and $f_1^R$ and $f_2^{L,R}$ are almost zero (SM values).

The measurement of $f_1^L$ can be interpreted as the measurement of the CKM element $V_{tb}$ which is therefore known to within 3%. (We note that this result is more accurate than that obtained from measuring the $qq' \to W^* \to t\bar{b}$ production rate which yields a 10% error in measuring $V_{tb}$ [1].) Furthermore, the fraction ($F_L$) of longitudinal $W$'s from top decays is equal to $\frac{1}{2}\frac{m_t^2}{M_W^2}$, and is independent of $f_1^{L,R}$, it is therefore a good tool for measuring $m_t$ [9]. (The quadratic dependence of $F_L$ on $m_t$ helps in $\delta m_t$ by a factor of 2.) A 2% accuracy in determining $F_L$ yields a $\sim 2\,\text{GeV}$ error in measuring $m_t$.

*3.8 Exotic production mechanism and testing CP violation in top*

If new physics strongly modifies the coupling of $t$-$c$-$g$, then the production rate of $t$-$c$ pair from $q\bar{q} \to g \to t\bar{c}$ can be largely enhanced. By measuring its production rate in high $M_{tc}$, one can set the minimum energy scale $\Lambda_{tcg}$ at which new physics must set in. It is found that $\Lambda_{tcg} \gtrsim 2\,\text{TeV}$ [13].

Besides all the potential physics discussed above, the Tevatron, as a $\bar{p}p$ collider, is unique for being able to test CP violation by measuring the production rates of single-top events. A nonvanishing asymmetry ($\mathcal{A}_t$) in the inclusive production rates of the single-$t$ events and the single-$\bar{t}$ events signals CP violation [14]. With $2\,\text{fb}^{-1}$, it is possible to observe this asymmetry for $\mathcal{A}_t \gtrsim 20\%$ [14].

This work is supported in part by the NSF under grant no. PHY-9309902. We thank our colleagues who contribute to the studies in Ref. [1].